\documentclass[%
 reprint,
 prd,
 amsmath,amssymb,
 aps,
nofootinbib
]{revtex4-2}

\usepackage{graphicx}
\usepackage{dcolumn}
\usepackage{bm}
\usepackage{mathtools}
\usepackage{xcolor}

\newcommand{\paren}[1]{\left(#1\right)}

\newcommand{\pfrac}[2]{\paren{\frac{#1}{#2}}}
\newcommand{\abs}[1]{\left|#1\right|}

\usepackage{url}
\usepackage[colorlinks=true, linkcolor=black, citecolor=black]{hyperref}
\usepackage{comment}
\usepackage{placeins}

\begin{document}


\title{Gravitational Wave Scattering From the Sun and Planets}

\author{Daniel A. Kessler}
\email{dak179@case.edu}
\author{Craig J. Copi}
\email{craig.copi@case.edu}
\author{Glenn D. Starkman}
\email{glenn.starkman@case.edu}
\affiliation{Department of Physics/CERCA/ISO, Case Western Reserve University, Cleveland OH 44106, USA}

\date{\today}

\begin{abstract}

General relativity predicts that massless waves should scatter from the Riemann curvature of their backgrounds. These scattered waves are sometimes called \textit{tails} and have never been directly observed. Here we calculate the gravitational waves scattered in the backward direction (scattering angle $\vartheta=\pi$) from the weak-field curvature of an extended massive object, finding close agreement with previous results in the long-wavelength limit. These long-wavelength results are then applied to gravitational waves in the Laser Interferometer Space Antenna (LISA) sensitivity band scattering from the Sun and planets. We estimate that scattering from the Sun and the planets from Jupiter to Neptune could contribute a $10^{-3}$ amplitude modulation to LISA observations when these objects almost intersect the line of sight between LISA and the source, leading to forward scattering with $\vartheta\simeq0$. These conditions should be realized for the Sun during the lifetime of LISA if the detection rate of long-duration sources is not much smaller than a thousand per year. 

\end{abstract}

\maketitle

\section{Introduction}

An early and prominent success of general relativity was the correct prediction of the gravitational bending of light \cite{Einstein1936}. It is less known that the theory predicts another change to massless wave propagation: such waves should scatter from the Riemann curvature of their backgrounds and be ``smeared" inside the lightcone \cite{Hadamard1923, DewittBrehme1960, DeWittDeWitt1964, Peters1974}. 
This is because all (commonly studied) massless fields couple to the Riemann tensor and its contractions once their equations of motion are made generally covariant \cite{MisnerWheelerThorne1973, PoissonPoundVega2011} and consistent with the equivalence principle \cite{SonegoFaraoni1993}. 

An interesting question is whether these scattered waves---sometimes called \textit{tails}---are directly observable. On this question, gravitational waves are an ideal candidate for study since they can scatter from regions of large Riemann curvature (\textit{e.g.}, inside stars) without being absorbed by the accompanying mass energy. Although the differential cross-section (DCS) of gravitational wave scattering in the weak-field and long-wavelength limits was derived by several authors almost fifty years ago \cite{Westervelt1971, Peters1976, LogiKovacs1977, Dolan2008, Guadagnini2008, Sorge2015}, the yet null observation of gravitational waves prevented much speculation on directly observing their scattering. In the modern era, such observations are a serious possibility and deserve more consideration.

Here we discuss the prospect of observing gravitational wave scattering in the solar system. To progress beyond the long-wavelength DCS results, we begin by outlining a straightforward, geometric method to obtain gravitational scattering amplitudes from the stress-energy tensor of the wave source and the density function of the scatterer. This method builds on \cite{DewittBrehme1960, PfenningPoisson2002, ChuStarkman2011, ChuPasmatsiouStarkman2020, CopiPasmatsiouStarkman2021, CopiStarkman2022} and in Section \ref{sec:methods} is illustrated using the simplest analytically tractable case of monotone gravitational waves scattered in the backward direction (scattering angle $\vartheta=\pi$) from the weak-field curvature of a spherically symmetric massive object with a smooth density profile. An analogous calculation was completed in \cite{CopiStarkman2022} for scattering in the forward direction, where $\vartheta\simeq0$. We show in Section \ref{sec:results} that, in both cases, the density function of the scatterer becomes irrelevant in the long-wavelength limit, and the resulting scattering amplitude closely agrees with the previous DCS results. In Section \ref{sec:scattering-sun-planets}, we discuss how gravitational waves in the LISA sensitivity band (centered around mHz frequencies \cite{LISA2024}) would scatter from the Sun and planets and how this scattering might be observed. These results and the probability of an observation are further discussed in Section \ref{sec:discussion}. Throughout, we use the geometrized units $c=G=1$ and Minkowski metric signature $(+,-,-,-)$.

\section{\label{sec:methods}Methods}

Our approach to the scattering calculation uses the Hadamard construction \cite{Hadamard1923} of Green's functions on weakly curved spacetimes, which was first studied physically by DeWitt and Brehme \cite{DewittBrehme1960} and DeWitt and DeWitt \cite{DeWittDeWitt1964} before the components of the gravitational wave Green's function were obtained explicitly by Pfenning and Poisson \cite{PfenningPoisson2002} and Chu and Starkman \cite{ChuStarkman2011} using different perturbative schemes. An important takeaway from these later studies is that, when spacetime is perturbed by nonrelativistic matter, the Green's function is entirely determined by the density function of the matter distribution. Once this function is specified, the scattered gravitational wave can be obtained by convolving the resulting Green's function with the stress-energy tensor of the wave source. Alternative approaches to such scattering calculations have used Feynman diagrams \cite{LogiKovacs1977, Guadagnini2008}, partial wave methods \cite{MatznerRyan1977, Dolan2008, StrattonDolan2019}, and analogies with optics \cite{Peters1974, Takahashi2005}, recently using a path integral formalism \cite{Braga:2024pik}.

In this section, we illustrate our methodology with the case of backward scattering, a compact and spherically symmetric density function, and the stress-energy tensor of a Newtonian binary system. The scattering geometry is described first in Section \ref{sec:background}, the gravitational wave perturbation in Section \ref{sec:gravitational-wave}, the wave source in Section \ref{sec:source}, and finally the Green's function in Section \ref{sec:greens-function}.

\subsection{\label{sec:background}Scattering Geometry}

\begin{figure}
    \centering
    \includegraphics[width=0.9\linewidth]{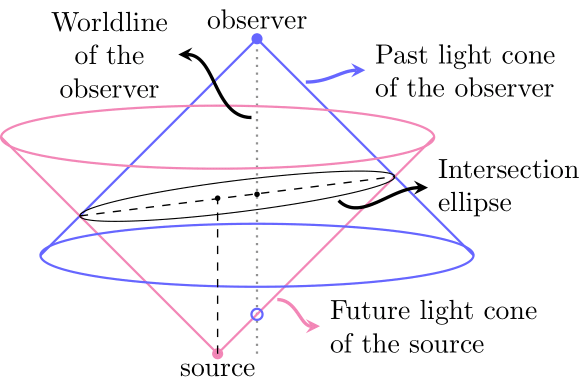}
    \caption{(Adapted from Figure 1 of \cite{CopiStarkman2022}.) Massless waves scatter from the curvature on the intersection of the source's future lightcone and observer's past lightcone. In three spatial dimensions and at a constant time, this intersection is an ellipsoid (in two spatial dimensions, it is an ellipse) with the source and observer locations as foci (black dots). The observer's worldline is shown (gray dotted line) alongside the arrival of the null wave at the observer (blue circle).}
    \label{fig:lightcones}
\end{figure}

The metric of the background spacetime is assumed to be
\begin{equation}
    \bar g_{\mu\nu}(\bm x)=\eta_{\mu\nu}+\delta\eta_{\mu\nu}(\bm x)\,,
    \label{eq:background-metric}
\end{equation}
where $\delta\eta_{\mu\nu}\ll\eta_{\mu\nu}$ is the static perturbation from an extended massive object, the \textit{perturber} of the spacetime,
whose mass $M$ and radius $a$ satisfy
\begin{equation}
    M\ll a\,.
    \label{eq:weak-field}
\end{equation}
The background spacetimes of moving perturbers can be approximated by Eq.\ \eqref{eq:background-metric} when treating massless wave scattering if the perturber moves much more slowly than light.\footnote{The background curvature depends on the d'Alembertian of the perturber's gravitational potential \cite{PfenningPoisson2002}. When the perturber's velocity is non-relativistic, the d'Alembertian can be approximated by the Laplacian.}

\begin{figure}
    \centering
    \includegraphics[width=0.95\linewidth]{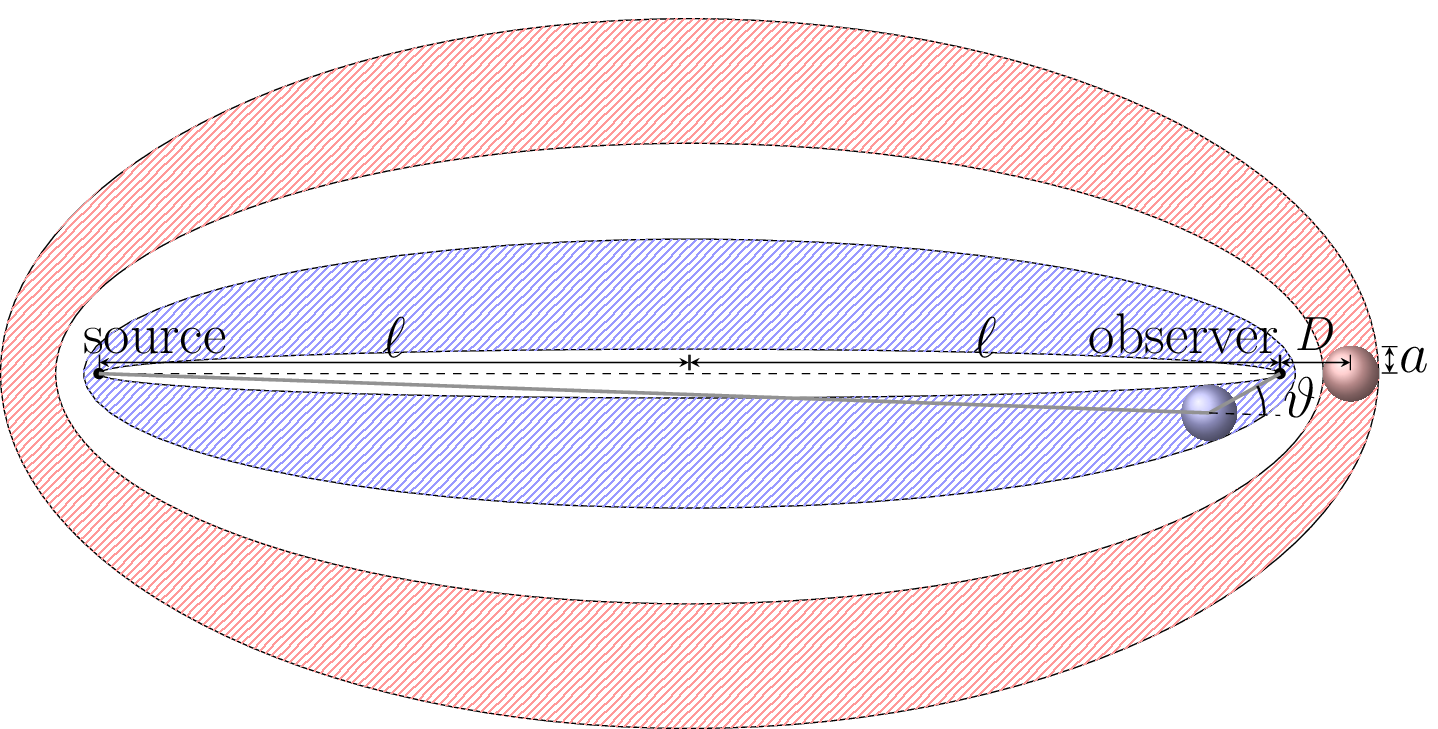}
    \caption{An illustration of the scattering geometry, defined by the observer-source distance $2\ell$, the observer-perturber distance $D$, the perturber radius $a$, and the scattering angle $\vartheta$. The ellipses (ellipsoid cross-sections) where the observed scattering occurs during the middle-time are shown for two possible locations of the perturber (blue and red shaded regions, corresponding to forward and backward scattering, respectively). To illustrate our methods, we specialize to backward scattering in Section \ref{sec:methods}. Forward scattering is the focus of our discussion in Section \ref{sec:scattering-sun-planets} on scattering from the Sun and planets.}
    \label{fig:scattering-geometry}
\end{figure}

An instrument observing a source of massless waves on a curved background will see the null signal followed by a scattering signal (sometimes called the \textit{tail}) that originates from the Riemann curvature on the intersection of its past lightcone and the source's future lightcone (Figure \ref{fig:lightcones}). 
The scattering geometry is defined by the distance $2\ell$ between the observer and source, the distance $D$ between the observer and perturber, the radius $a$ of the perturber, and the scattering angle $\vartheta$ (Figure \ref{fig:scattering-geometry}). As the simplest case that applies to scattering in the solar system, we specialize to the backward $\vartheta=\pi$ direction and assume the inequalities
\begin{equation}
    a\ll D\ll \ell\,,
    \label{eq:geometry-inequalities}
\end{equation}
which hold when the observer is LISA, the perturber is the Sun or planets, and the source is far outside the solar system.

The semi-major axis $s$ of the ellipsoid on which the observed scattering occurs can be written
\begin{equation}
    s=\frac{t-t'}{2}\,,
\end{equation}
where $t$ is the observation time and $t'$ is the source emission time \cite{PfenningPoisson2002}. The scattering signal is largest when this ellipsoid intersects the interior of the perturber, where the background is most curved, and the corresponding range of times (or time differences) has been called the \textit{middle-time} \cite{CopiPasmatsiouStarkman2021}. It was shown in \cite{CopiPasmatsiouStarkman2021, CopiStarkman2022} that the dominant part of the middle-time signal originates from the interior of the perturber.
An observer hoping to measure scattering in the solar system should hence ``point" their detector toward the Sun and planets. Our focus in this paper is predicting the middle-time, interior signal they would see. 

For scattering in the backward direction, the middle-time corresponds to
\begin{equation}
    s\in[s_0-a,s_0+a]\,,
\end{equation}
where $s_0$ is the semi-major axis of the ellipsoid (also focused at the observer and source) that intersects the center of the perturber. The corresponding semi-minor axis $\gamma_0=\sqrt{s_0^2-\ell^2}$ will make an appearance in Section \ref{sec:results}.


\subsection{\label{sec:gravitational-wave}Gravitation Wave}

The full metric of the spacetime is 
\begin{equation}
    g_{\mu\nu}(x)=\bar g_{\mu\nu}(\bm x)+h_{\mu\nu}(x)\,,
\end{equation}
where $h_{\mu\nu}\ll1$ is the perturbation from a gravitational wave propagating over the background defined by $\bar g$ in Eq.\ \eqref{eq:background-metric}. The trace-reversed perturbation
\begin{equation}
    \tilde h_{\mu\nu}=  h_{\mu\nu}-\frac{1}{2}\big(\bar g^{\rho\sigma}h_{\rho\sigma}\big)\bar g_{\mu\nu}
\end{equation}
in the Lorenz gauge $\bar\nabla^\mu\tilde h_{\mu\nu}=0$ obeys the wave equation (see for example \cite{MisnerWheelerThorne1973})
\begin{equation}
    \bar g^{\rho\sigma}\bar\nabla_\rho\bar\nabla_\sigma\tilde h_{\mu\nu}+2 \bar R_{\rho\mu\sigma\nu}\tilde h^{\rho\sigma}=16\pi \delta T_{\mu\nu}\,,
\end{equation}
where $\bar R_{\rho\mu\sigma\nu}$ is the Riemann tensor of the background, $\bar\nabla$ is the background covariant derivative, and $\delta T_{\mu\nu}$ is the stress-energy tensor of the gravitational wave source. This equation can be solved using the trace-reversed Green's function $\tilde G_{\mu\nu\alpha'\beta'}$, after which one can trace-reverse again to obtain $ G_{\mu\nu\alpha'\beta'}$ \cite{PfenningPoisson2002}. The perturbation $h_{\mu\nu}$ then follows from the convolution
\begin{equation}
    h_{\mu\nu}(x)\simeq 4\int d^4x'\:G_{\mu\nu\alpha'\beta'}(x,x')\,\delta T^{\alpha'\beta'}(x')\,,
    \label{eq:h-convolution}
\end{equation}
where primed indices are associated with $x'$.

\subsection{\label{sec:source}Source}

We assume that the gravitational wave source is a Newtonian (slow-moving and well-separated) binary of equal masses $M_b$ in a mutual circular orbit of radius $R_b$, hence orbiting with velocity
\begin{equation}
    \Omega_b=\sqrt{\frac{M_b}{4R_b^3}}\,.
\end{equation}

Using the known expression for the stress-energy tensor of point masses (see for example \cite{Weinberg1972}), we find for the Newtonian binary
\begin{gather}
    \delta T_{\mu\nu}(x)\simeq M_b\bigg\{\delta^3[\bm x-\bm x_b^{+}(t)]+\delta^3[\bm x-\bm x_b^{-}(t)]\bigg\}\delta\hat T_{\mu\nu}(\bm x)\,,\nonumber\\
    \delta\hat T_{\mu\nu}(\bm x)=
    \begin{pmatrix}
    1 & -\Omega_b y & \Omega_b x & 0 \\
    -\Omega_b y & \Omega_b^2 y^2 & -\Omega_b^2 x y & 0 \\
    \Omega_b x & -\Omega_b^2 xy & \Omega_b^2x^2 & 0 \\
    0 & 0 & 0 & 0
    \end{pmatrix}\,,
    \label{eq:newtonian-binary}
\end{gather}
where the binary objects' positions are denoted
\begin{equation}
    \bm x_b^\pm(t)=\big(\pm R_b\cos\Omega_b t,\:\pm R_b\sin\Omega_b t,\:z_b\big)\,.
\end{equation}
This stress-energy tensor includes time-independent contributions from the binary's mass and kinetic energy, which need to be removed when calculating the gravitational wave perturbation.

\subsection{\label{sec:greens-function}Green's Function}

When spacetime is sufficiently weakly curved that there is a unique geodesic between the observer at $x$ and source at $x'$, the Green's function in Eq.\ \eqref{eq:h-convolution} can be written in the Hadamard form
\begin{multline}
    G_{\mu\nu\alpha'\beta'}(x,x')=G^\mathrm{null}_{\mu\nu\alpha'\beta'}(x,x')\delta(\sigma)\\+G^\mathrm{scat}_{\mu\nu\alpha'\beta'}(x,x')\theta(-\sigma)\,,
    \label{eq:hadamard-form}
\end{multline}
where $\sigma$ is Synge's world function \cite{Hadamard1923,PoissonPoundVega2011}.\footnote{See Sections 3 and 16.2 of the review article \cite{PoissonPoundVega2011} for more details.} In Eq.\ \eqref{eq:hadamard-form}, the first term describes the propagation of the null signal on the lightcone while the second term describes the scattering signal inside the lightcone. The scattering Green's function has been calculated by several authors using different perturbative schemes \cite{PfenningPoisson2002, ChuStarkman2011}. To lowest order in the background perturbation, its components are
\begin{align}
		G^\mathrm{scat}_{tttt} &= -\partial_{tt'}A+B \nonumber\\
	G^\mathrm{scat}_{ttta} &= - G^\mathrm{scat}_{tatt}=\paren{\partial_{t'a}-\partial_{ta'}}A \nonumber\\
     G^\mathrm{scat}_{abtc} &= -G^\mathrm{scat}_{tcab}=\delta_{c(a}\paren{\partial_{b)t'}-\partial_{b')t}}A \nonumber\\
	G^\mathrm{scat}_{ttab} &= G^\mathrm{scat}_{abtt}=\paren{\partial_{a}+\partial_{a'}}\paren{\partial_{b}+\partial_{b'}}A-\delta_{ab}\paren{\partial_{tt'}A-B} \nonumber\\
     G^\mathrm{scat}_{tatb} &= \delta_{ab}\paren{\partial_{tt'}A-B}-\frac{1}{2}\paren{\partial_{ab}+2\partial_{a'b}+\partial_{a'b'}}A \nonumber\\
	G^\mathrm{scat}_{abcd} &=\paren{2{\delta^{(a}}_{c}{\delta^{b)}}_{d}-\delta_{ab}\delta_{cd}}
	\paren{-\partial_{tt'}A} \nonumber\\ &+\delta_{ab}\paren{\partial_{c}+\partial_{c'}}\paren{\partial_{d}+\partial_{d'}}A  \nonumber\\   &-2{\delta^{(a}}_{(c}\paren{{\partial^{b)}}_{d)}+2{\partial^{b)}}_{d')}+{\partial^{b')}}_{d')}}A \nonumber\\ &+\delta_{cd}\paren{\partial_{a}+\partial_{a'}}\paren{\partial_{b}+\partial_{b'}}A 
    +\delta_{ab}\delta_{cd} B\,,
    \label{eq:scattered-Green's}
\end{align}
where spatial indices are represented by Latin letters and we have suppressed the primes in $G^\mathrm{scat}_{\mu\nu\alpha'\beta'}$ for readability. To further simplify these expressions, we have used the shorthand from \cite{PfenningPoisson2002}
\begin{gather}
    \partial_{\mu\nu}\equiv \partial_\mu\partial_\nu\,,\:u_{(a}v_{b)}\equiv \frac{1}{2}\big(u_av_b+u_bv_a\big)\,,\:B\equiv {\partial^\rho}_{\rho'}A\,.
\end{gather}
Every component of the scattering Green's function thus equals a second-order differential operator acting on the $A$ function,
\begin{multline}
    A(x,x')=\frac{1}{4\pi}\int d\Omega\, \Phi\bigg(\bm{r}_\mathrm{ellipsoid}(s,\theta,\phi)\\-\bm{r}_\mathrm{perturber}(s_0,\theta_0,\phi_0)\bigg),
    \label{eq:A-function-definition}
\end{multline}
which represents the average background gravitational potential $\Phi$ over the intersection of the observer and source lightcones \cite{PfenningPoisson2002}. In Eq.\ \eqref{eq:A-function-definition}, $\bm r_\mathrm{ellipsoid}$ ranges over this intersection while $\bm r_\mathrm{perturber}$ is fixed at the center of the perturber. The gravitational potential is obtained from the Poisson equation\footnote{This convention agrees with \cite{PfenningPoisson2002, PoissonPoundVega2011} but differs from \cite{ChuStarkman2011, CopiStarkman2022} by a factor of $4\pi$.}
\begin{equation}
    \Phi(\bm{r})=-\int d^3\bm{r}'\frac{\rho(\bm{r}')}{\left|\bm{r}-\bm{r}'\right|}\,,
    \label{eq:gravitational-potential}
\end{equation}
where for simplicity we assume that the mass density $\rho$ inside the perturber takes the smooth polynomial form
\begin{equation}
    \rho(r)= \rho_\text{central}\left( 1 - \frac{r^2}{a^2} \right)^p\,.
    \label{eq:mass-density}
\end{equation}
The exponent $p$ is an integer that should be large enough (\textit{i.e.}, make the $A$ function differentiable enough) so that observable quantities are continuous. We find that $p=4$ is sufficient for $\ddot h$. The \textit{A} function corresponding to Eq.\ \eqref{eq:mass-density} for scattering in the backward direction and general $p$ is derived in Appendix \ref{sec:A-function}. 

\section{\label{sec:results}Results}

The scattered part $h_{\mu\nu}^\mathrm{scat}$ of the gravitational wave perturbation in Eq.\ \eqref{eq:h-convolution} can now be calculated using the stress-energy tensor in Eq.\ \eqref{eq:newtonian-binary} and scattering Green's function in Eq.\ \eqref{eq:scattered-Green's}, assuming the density function in Eq.\ \eqref{eq:mass-density}. We present the results in the long-wavelength limit, where the gravitational wave frequency $2\Omega_b$ satisfies
\begin{equation}
    a\ll (2\Omega_b)^{-1}\ll\ell\,.
    \label{eq:long-wavelength}
\end{equation}

To lowest-order in the small quantities created by Eqs.\ \eqref{eq:weak-field}, \eqref{eq:geometry-inequalities}, and \eqref{eq:long-wavelength}, the scattered perturbation is transverse and traceless with transverse components
\begin{multline}
    h^\mathrm{scat}_{ij}\simeq \left[2M_b(R_b\Omega_b)^2\right]\,\frac{2M}{\gamma_0^2}\:\times\\
    \times\begin{pmatrix}
        -\cos2\Omega_bt_r^\mathrm{scat} & -\sin2\Omega_bt_r^\mathrm{scat} \\
        -\sin2\Omega_bt_r^\mathrm{scat} & \cos2\Omega_bt_r^\mathrm{scat}
    \end{pmatrix} \,,
    \label{eq:h-scat}
\end{multline}
where $t_r^\mathrm{scat}=t-2s_0$ and $\gamma_0$ is the semi-minor axis mentioned in Section \ref{sec:background}.
The null wave at the observer is approximately the same as it is in flat spacetime,
\begin{multline}
    h^\mathrm{null}_{ij}\simeq \left[2M_b(R_b\Omega_b)^2\right]\,\frac{2}{\ell}\:\times \\ 
    \times \begin{pmatrix}
        -\cos2\Omega_bt_r^\mathrm{null} & -\sin2\Omega_bt_r^\mathrm{null} \\
        -\sin2\Omega_bt_r^\mathrm{null} & \cos2\Omega_bt_r^\mathrm{null}
    \end{pmatrix} \,,
    \label{eq:h-null}
\end{multline}
where $t_r^\mathrm{null}=t-2\ell$ (see for example \cite{Carroll2019}). The ratio $\mathcal R$ of the scattering and null signal amplitudes at the observer is then
\begin{equation}
    \mathcal R\simeq\frac{M\ell}{\gamma_0^2}\simeq\frac{M}{2D}\,.
    \label{eq:amplitude-ratio}
\end{equation}
Since $h_{ij}^\mathrm{scat}$ and $h_{ij}^\mathrm{null}$ have the same time-dependence, Eq.\ \eqref{eq:amplitude-ratio} also gives the ratio of the observed Riemann curvature $R_{0i0j}\propto\ddot h$ and square-rooted energy flux $\sqrt{F}\sim\dot h$ from the scattered and null gravitational waves. It is remarkable that, in the long-wavelength limit, the observed amplitude ratio depends only on the perturber mass $M$ and distance $D$ from the observer.

\subsection{\label{sec:dcs}General Scattering Angle}

Scattering processes are traditionally characterized by the fraction of incident energy flux scattered into a solid angle element $d\Omega$, defined as the differential cross-section $d\sigma/d\Omega$. From this, one obtains the scattering amplitude
\begin{equation}
    \mathcal A=\sqrt{\frac{d\sigma}{d\Omega}}\,,
\end{equation}
which is subtly different than the amplitude ratio $\mathcal R$ in Eq.\ \eqref{eq:amplitude-ratio}. The scattering amplitude implied by that equation is
\begin{equation}
    \mathcal A=\frac{2\ell}{2\ell+D}\,D\,\mathcal{R}\simeq\frac{M}{2}\,.
    \label{eq:implied-scattering-amplitude}
\end{equation}
Previous authors \cite{Westervelt1971, Peters1976, LogiKovacs1977, Dolan2008, Guadagnini2008, Sorge2015} have obtained the differential cross-section
\begin{equation}
    \frac{d\sigma}{d\Omega}\simeq M^2\left[\frac{\cos^2\vartheta+(1/8)\sin^4\vartheta}{\sin^4(\vartheta/2)}\right]\,
    \label{eq:differential-cs}
\end{equation}
for unpolarized gravitational waves scattered at an angle $\vartheta$ from the Riemann curvature of a point-like or extended mass $M$ in the limit of weak gravitational fields and long incident wavelengths. 
Equation \eqref{eq:differential-cs} implies the backward $\vartheta=\pi$ and forward $\vartheta\simeq0$ direction scattering amplitudes
\begin{equation}
    \mathcal A(\vartheta=\pi)\simeq M\,,\quad \mathcal A(\vartheta\simeq0)\simeq\frac{4 M}{\vartheta^2}\,,
    \label{eq:scattering-amplitudes}
\end{equation}
which are factors of two larger than the one we derived in Eq.\ \eqref{eq:implied-scattering-amplitude} for the backward direction and the one derived in \cite{CopiStarkman2022} using a similar methodology for the forward direction.\footnote{In \cite{CopiStarkman2022}, the scattering amplitude implied by their Eq.\ (10) is
\begin{equation}
    \mathcal A\simeq\frac{\ell}{2\ell}\,\ell\left(\frac{4M}{\ell\vartheta^2}\right)=\frac{2M}{\vartheta^2}\,.
\end{equation}
} 
For definiteness, we will use the expressions in Eq.\ \eqref{eq:scattering-amplitudes} for the calculations to follow. The observed amplitude ratios analogous to Eq.\ \eqref{eq:amplitude-ratio} implied by the scattering amplitudes in Eq.\ \eqref{eq:scattering-amplitudes} are
\begin{equation}
    \mathcal R(\vartheta=\pi)\simeq \frac{M}{D}\,,\quad \mathcal R(\vartheta\simeq0)\simeq\frac{4 M}{D\vartheta^2}\,.
    \label{eq:implied-amplitude-ratios}
\end{equation}
As a consequence of taking the long-wavelength limit, these expressions effectively assume that the scatterer is a point mass. Hence, although the forward scattering ratio diverges as the scattering angle approaches zero, the smallest angle supported for the extended scatterers we consider is $\vartheta\simeq a/D$.\footnote{The calculation in \cite{CopiStarkman2022} also prohibits smaller scattering angles. When the perturber eclipses the line of sight, it intersects the scattering ellipsoid in two separate regions and the middle-time, interior \textit{A} function must account for this.}

\subsection{\label{sec:scattering-sun-planets}Scattering from the Sun and Planets}

We use the forward scattering ratio in Eq.\ \eqref{eq:implied-amplitude-ratios} to estimate the largest observable consequences of gravitational wave scattering in the solar system. The long-wavelength condition [Eq.\ \eqref{eq:long-wavelength}] restricts us to gravitational waves with frequencies below $10^{-2}$ Hz, which falls within the expected sensitivity band of LISA \cite{LISA2024}. Hereafter, we consider LISA the observer and tailor our discussion to its parameters.

\subsubsection{Directional Observation}

\begin{figure}
    \centering
    \includegraphics[width=0.95\linewidth]{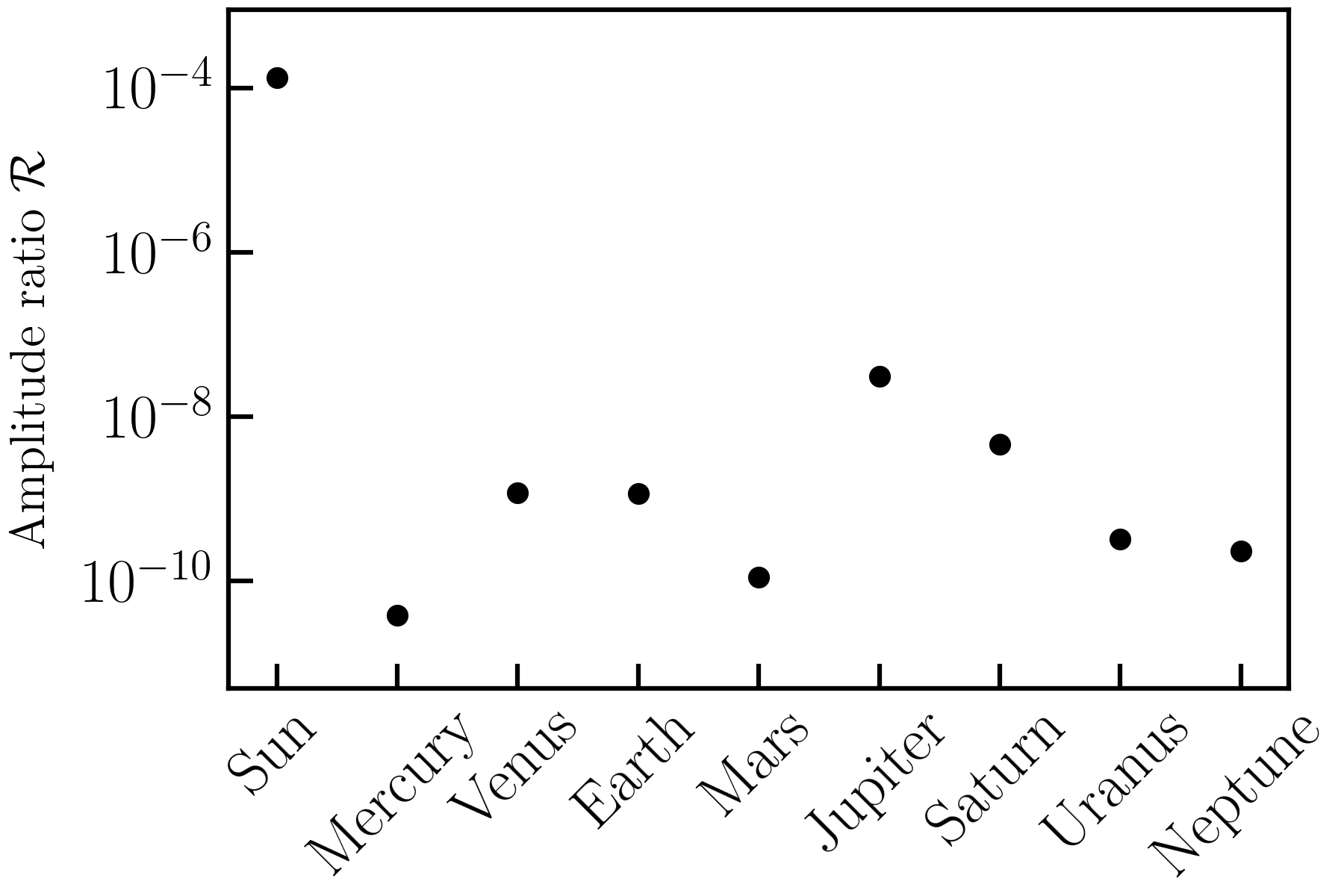}
    \caption{The largest directionally resolvable amplitude ratio $\mathcal R$ [Eq.\ \eqref{eq:resolvable-R}] for scattering from the Sun and planets, assuming that the angular resolution of LISA is $\theta_\mathrm{LISA}=1^\circ$. The observer-perturber distance $D$ is assumed to be the minimal distance between the perturber and LISA \cite{Arlot, Williams}, which is expected to launch to an Earth-like orbit around the Sun, trailing the Earth by approximately 50 million kilometers \cite{LISA2024}.} 
    \label{fig:resolvable-amplitude-ratio}
\end{figure}

\begin{figure}
    \centering
    \includegraphics[width=0.95\linewidth]{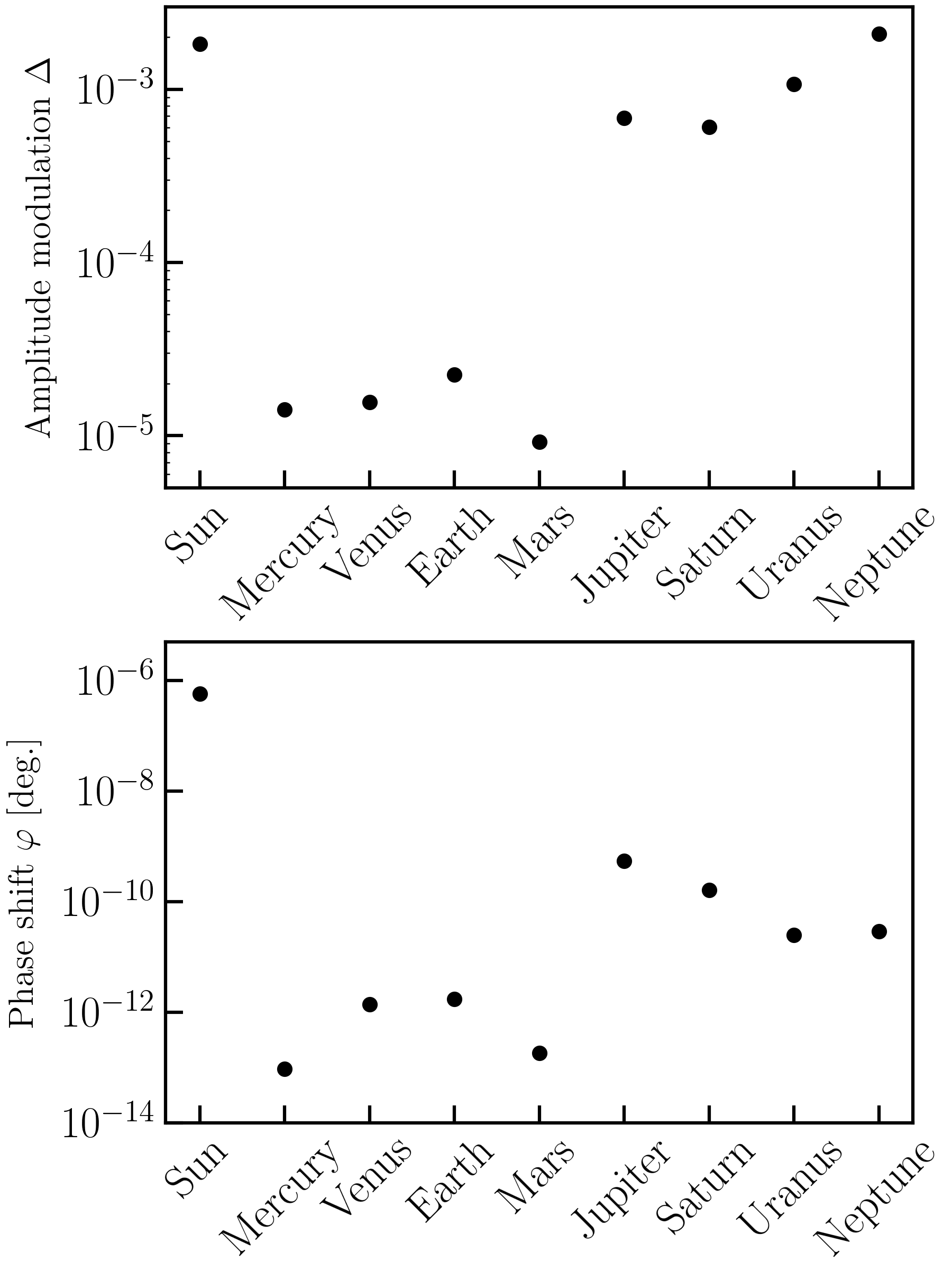}
    \caption{The largest amplitude modulations and the corresponding phase shifts [Eqs.\ \eqref{eq:ampl-mod-phase-shift} and \eqref{eq:maximum-ampl-mod}] caused by scattering from the Sun and planets, assuming the incident gravitational wave has frequency $2\Omega_b=1$ mHz. See the caption of Figure \ref{fig:resolvable-amplitude-ratio} for a remark on the observer-perturber distances assumed. The planets from Jupiter to Neptune can contribute as large an amplitude modulation as the Sun since these objects permit smaller scattering angles without violating the condition $\vartheta\gtrsim a/D$.}
    \label{fig:ampl-mod-phase-shift}
\end{figure}

For the direction of the scattered wave to be distinguishable from the null wave, it must be separated from the line of sight between LISA and the source by more than the angular resolution $\theta_\mathrm{LISA}$ of the detector. The largest resolvable amplitude ratio is then
\begin{multline}
    \mathcal R\simeq 1.3\times10^{-4}\:\left(\frac{M}{M_\odot}\right)\left(\frac{1\text{ AU}}{D}\right)\left(\frac{1^\circ}{\theta_\mathrm{LISA}}\right)^2\,.
    \label{eq:resolvable-R}
\end{multline}
This amplitude ratio is plotted for the Sun and planets in Figure \ref{fig:resolvable-amplitude-ratio} assuming $\theta_\mathrm{LISA}=1^\circ$, as motivated by optimistic estimates in \cite{Cutler1998}. Even for scattering from the Sun, where $\mathcal R$ is largest, the scattering signal would likely be difficult to observe on its own. An effort to make this observation might be assisted by the fact that the scattered wave has the same frequency as the null wave and is delayed by a known amount,
\begin{equation}
    t_\mathrm{delay}\simeq\frac{D\vartheta^2}{2}\simeq 8\times10^{-2}\text{ seconds}\: \left(\frac{D}{1\text{ AU}}\right)\left(\frac{\vartheta}{1^\circ}\right)^2\,.
    \label{eq:t-delay}
\end{equation}

\subsubsection{Amplitude Modulation}

\begin{figure}
    \centering
    \includegraphics[width=0.95\linewidth]{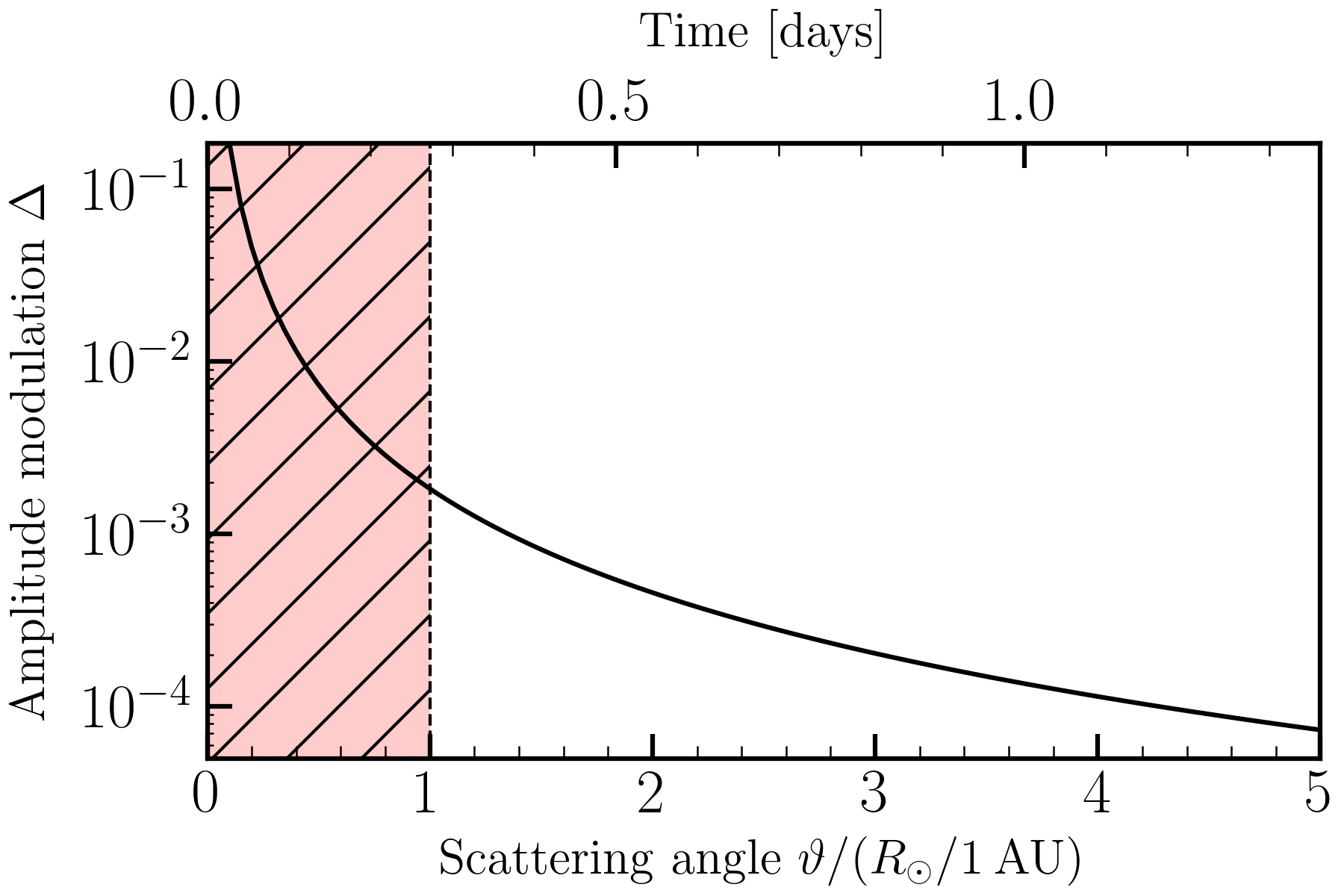}
    \caption{The amplitude modulation to LISA observations caused by scattering from the Sun. The red-hatched region corresponds to the Sun eclipsing the line of sight between LISA and the source, where our formalism breaks down. The top axis measures the time since total eclipse and shows that the amplitude modulations supported by our formalism, beginning when the Sun is just outside the line of sight, change by around an order of magnitude over the course of a day. The corresponding phase shift $\varphi$ remains on the order of $10^{-6}$ during this period.}
    \label{fig:ampl-mod-theta}
\end{figure}

If the direction of the scattered wave cannot be resolved, the scattering signal could still be observed through the amplitude modulation and phase shift it contributes to the overall signal. 
In this case, the detector observes the sum
\begin{multline}
    h_{ij}^\mathrm{null}+h_{ij}^\mathrm{scat}\sim h^\mathrm{null}\left[\cos2\Omega_bt_r^\mathrm{null}+\mathcal{R}\cos2\Omega_bt_r^\mathrm{scat}\right] \\
    \equiv h^\mathrm{null}(1+\Delta)\cos\left(2\Omega_bt_r^\mathrm{null}-\varphi\right)
    \,,
\end{multline}
where $h^\mathrm{null}$ is the maximum amplitude of the null wave. For scattering angles much smaller than unity, the amplitude modulation $\Delta$ and phase shift $\varphi$ are approximately
\begin{equation}
    \Delta\simeq\mathcal R\,,\quad\varphi\simeq\frac{\mathcal R }{1+\mathcal R}\,\Omega_b D\,\vartheta^2\,.
    \label{eq:ampl-mod-phase-shift}
\end{equation}
The bound $\vartheta\gtrsim a/D$ on the forward scattering amplitude ratio $\mathcal{R}$ limits the amplitude modulation to
\begin{multline}
    \Delta\lesssim 4\left(\frac{M}{a}\right)\left(\frac{D}{a}\right)\\\simeq 1.8\times10^{-3}\:\left(\frac{M}{M_\odot}\right)\left(\frac{D}{1\text{ AU}}\right)\left(\frac{R_\odot}{a}\right)^2 .
    \label{eq:maximum-ampl-mod}
\end{multline}
This limit and the corresponding phase shifts are plotted for the Sun and planets in Figure \ref{fig:ampl-mod-phase-shift}. If the scattering geometry was time-independent and the arrival delay [Eq.\ \eqref{eq:t-delay}] of the scattered wave could not be resolved, then the scattering signal would only be observable given detailed information about the source's intrinsic amplitude and distance from LISA. However, as the Sun moves across the sky at approximately one degree per day, the overall signal amplitude will slowly and perhaps observably migrate from the null signal amplitude (Figure \ref{fig:ampl-mod-theta}). Our chances of observing the maximal amplitude modulation from the Sun are briefly discussed in the following section.

\section{\label{sec:discussion}Discussion}

In this paper, we demonstrated a straightforward method to calculate gravitational wave scattering from extended massive objects that closely agrees with previous DCS results in the long-wavelength limit. These long-wavelength results were then applied to gravitational waves in the LISA sensitivity band scattering from the Sun and planets and two methods of observation were discussed: directional observation, when the perturber is separated from the line of sight between LISA and the source by more than the angular resolution of the detector, and amplitude modulation when the perturber is closer but does not intersect the line of sight. This restriction is necessary because the long-wavelength results effectively treat the scatterer as a point mass and diverge as the scattering angle approaches zero. Since any perturber approaching the line of sight so closely will probably intersect it, the eclipsing case should be further studied, and we defer this to future work.

The Sun or planets must be closely aligned with the observer and source for the resulting scattering to possibly be observable, at least in the near term. For the sake of discussion, assume an observation is possible when the center of the perturber is separated from the observer-source line by an angle $\alpha\in[\alpha_1,\alpha_2]$ in both the polar and azimuthal directions. A simple estimate for the observation probability is then $p_\alpha^2$ where $p_\alpha=(\alpha_2-\alpha_1)/\pi$. However, for sources with durations longer than a year and scattering from the Sun, the probability increases to $p_\alpha$ since the source only needs to be within $\alpha$ of the orbital plane of LISA for the required alignment to occur eventually. In this case, an observable amplitude modulation would result from approximately $(R_\odot/1\text{ AU})/\pi\sim10^{-3}$ of all sources and should occur over the lifetime of LISA if the detection rate of long-duration sources is not much smaller than a thousand per year.

This is not the first proposal for observing gravitational wave scattering. Reference \cite{CopiStarkman2022} for example studied finite duration gravitational waves with frequencies in the Laser Interferometer Gravitational-wave Observatory (LIGO) sensitivity band scattering from compact objects at cosmological distances from the observer (and from the source), finding that the scattering signal would ``echo" the original waveform with amplitude ratios as large as one-third. Furthermore, the effects of gravitational wave scattering could be (indirectly) observable in the waveforms themselves, as such scattering has been shown to alter the motion of binary systems through self-forces and higher-order corrections \cite{Blanchet2002, BarackPound2018}. In both of those cases, however, the scattering process occurs very far away under uncertain conditions. Our results suggest the solar system could be a more ``controlled" laboratory for testing the scattering predictions of general relativity.

\begin{acknowledgments}
We thank Madeline Wade and David Singer for helpful discussions in the early stages of this work. D.K. and G.D.S.\ acknowledge partial support from  DOE grant DESC0009946.
\end{acknowledgments}

\appendix

\section{\label{sec:A-function}$A$ Function}

Here we calculate the $A$ function defined in Eq.\ \eqref{eq:A-function-definition} for the mass density in Eq.\ \eqref{eq:mass-density} and scattering in the backward direction. The gravitational potential in Eq.\ \eqref{eq:gravitational-potential} is found to be
\begin{equation}
    \Phi(r)=
    \begin{dcases}
        -n_p\frac{M}{a}\sum_{k=0}^{p+1}\Phi^p_{k}\pfrac{r}{a}^{2k} & r\leq a \\
        -\frac{M}{r} & r>a\,,
    \end{dcases}
\end{equation}
where we have defined the coefficients
\begin{gather}
    \Phi^p_0\equiv1\,,\quad \Phi^p_{k>0}\equiv\frac{(-1)^k}{c_p}
    \begin{pmatrix}
        p \\ k - 1
    \end{pmatrix}
    \frac{1}{2k(2k+1)}\\
    c_p\equiv\sum_{k=0}^p(-1)^k\begin{pmatrix} p \\ k \end{pmatrix}\frac{1}{2k+2}
\end{gather}
and the function
\begin{equation}
    n_p\equiv\frac{\Phi(0)}{-M/a}
\end{equation}
that determines the central gravitational potential.
The $A$ function takes the piecewise form
\begin{equation}
    A(x,x')=\begin{dcases}
        A_\text{early} & s\in(0,\:s_0-a) \\
        A_\text{middle} & s\in[s_0-a,\:s_0+a] \\
        A_\text{late} & s\in(s_0+a,\:\infty)\,,
    \end{dcases}
\end{equation}
whose early and late-time pieces are well-known \cite{DeWittDeWitt1964, PfenningPoisson2002},
\begin{equation}
    A_\text{early}=-\frac{M}{2\ell}\ln\pfrac{s_0+\ell}{s_0-\ell},\: A_\text{late}=-\frac{M}{2\ell}\ln\pfrac{s+\ell}{s-\ell}\,.
\end{equation}
The contributions to the middle-time $A$ function
\begin{equation}
    A_\text{middle}=A_\text{interior}+A_\text{exterior}
\end{equation}
from inside and outside the perturber are found by requiring $\abs{\bm r_\mathrm{ellipse}-\bm r_\mathrm{perturber}}\leq a$ and $\abs{\bm r_\mathrm{ellipse}-\bm r_\mathrm{perturber}}>a$, respectively, in Eq.\ \eqref{eq:A-function-definition}. The results are
\begin{align}
        A_\text{interior} &= -\frac{n_p}{2}\frac{M}{a}\sum_{k=0}^{p+1}\frac{1}{2k+1}\frac{1}{\epsilon^{2k}}C_k^p\Delta_{k+1/2}
        \label{eq:A-interior}
        \\
        A_\text{exterior} &= -\frac{M}{2\ell}\ln\left[\frac{\paren{\Sigma^2+\delta_{s,+}^2}^{1/2}+\delta_{s,+}}{\paren{\Sigma^2+\epsilon^2}^{1/2}+\epsilon}\right],
    \label{eq:A-exterior}
\end{align}
where we have defined
\begin{multline}
    \Delta_{j}\equiv\paren{\Sigma^2+\epsilon^2}^j-\paren{\Sigma^2+\delta_{s,-}^2}^j \\
    C_k^p \equiv(-1)^k\sum_{j=0}^{p+1-k}\abs{\Phi_{k+j}^p}
    \begin{pmatrix}
		k+j \\ k
    \end{pmatrix}\pfrac{\Sigma}{\epsilon}^{2j}\\
    \Sigma\equiv\frac{\gamma\gamma_0}{\ell^2}\quad
    \epsilon\equiv \frac{a}{\ell}\quad 
     \delta_{s,\pm}\equiv\frac{s\pm s_0}{\ell}\,.
\end{multline}
Here, $\gamma=\sqrt{s^2-\ell^2}$ is the semi-minor axis of the ellipsoid with semi-major axis $s$, and $\gamma_0$ is analogous for $s_0$.


\bibliography{main}

\end{document}